\documentclass[12pt,preprint]{aastex}

\DeclareGraphicsExtensions{.jpg, .eps}
\slugcomment{Accepted to ApJ}

\shorttitle{Mid-Infrared Imaging of the Herbig Ae Star AB Aur}
\shortauthors{Marinas et al.}

\begin{document}

\title{Mid-Infrared Imaging of the Herbig Ae Star AB Aurigae: \\
    Extended Emission on Several Scales\altaffilmark{1}}

\author{N. Mari\~{n}as\altaffilmark{2}, C. M. Telesco\altaffilmark{2}, R. S. Fisher\altaffilmark {3}, C. Packham\altaffilmark{2} and J. T. Radomski\altaffilmark{2,4}}

\altaffiltext{1}{Based on observations obtained at the Gemini Observatory, which is operated by the Association of Universities for Research in Astronomy, Inc., under
a cooperative agreement with the NSF on behalf of the Gemini partnership: the National Science Foundation (United States), the Particle Physics and Astronomy Research
Council (United Kingdom), the National Research Council (Canada), CONICYT (Chile), the Australian Research Council (Australia), CNPq (Brazil) and CONICET (Argentina)}
\altaffiltext{2}{Astronomy Department, University of Florida,Gainesville, FL 32611}
\altaffiltext{3}{Gemini Observatory Northern Operations Center, Hilo, Hawaii 96720}
\altaffiltext{4}{Gemini Observatory Southern Operations Center, La Serena, Chile}

\begin{abstract}
We present high sensitivity sub-arcsecond resolution images of the Herbig Ae star AB Aurigae at 11.6 and 18.5 $\mu$m taken with Michelle on Gemini North.
Bright extended dust emission close to the star is resolved at both wavelengths, with quadratically subtracted FWHM of 17$\pm$4 AU at 11.6 $\mu$m and 22$\pm$5 AU at 18.5 $\mu$m. Additional, fainter emission is detected out to a radius of 280 AU at 11.6 $\mu$m and 350 AU at 18.5 $\mu$m down to the sensitivity limit of the observations. The latter value is identical to the measured size of the millimeter-continuum disk, but much smaller than the CO disk. Assuming moderately absorbing material, we find that larger particles ($\sim 1$ $\mu$m) dominate the mid-IR emission in the inner ($<$ 100 AU) regions of the disk, and smaller particles ($<$ 0.3 $\mu$m) dominate in the outer regions of the disk.  A model of a nearly face-on passive flared disk with an inner rim accounts well for our observations.
\end{abstract}


\section{Introduction}

Located only 144 pc away \citep{anc98}, the 2 Myr old A0 star AB Aurigae is the brightest (V=7.06) of the original sample of Herbig stars \citep{her60}, which are intermediate mass (2 to 8 M$\sun$) pre-main sequence stars. Consequently, it is not only the best-studied Herbig object, but it has become an important touchstone for our understanding of the class. The spectral energy distribution (SED) of this source shows emission in excess of the photosphere throughout the infrared region indicative of circumstellar (CS) dust. Different models, among which are a highly inclined passive flared disk with an inner rim \citep{dul01}; a flat thick disk surrounded by a halo \citep{vin03}; and a halo alone \citep{eli04}, have been used to explain the spatial distribution of this dust. Despite their differences, all of these models reproduce reasonably well the observed SED, which indicates the need for high-resolution imaging to provide additional critical constraints.

Spatial observations at various wavelengths imply that the CS dust in the AB Aur system lies in a disk and some type of more extended structure. An inhomogeneous envelope extending to 1300 AU is apparent in optical scattered light \citep{gra99}, while closer to the star, in optical and near-IR scattered light, one sees what appears to be a disk with quasi-spiral structure, a radius of 580 AU, and an inclination of 30$^\circ$ (face-on = 0$^\circ$), assuming flat geometry \citep{gra99, fuk04}.  CO observations reveal a complex disk with an inner hole of about 70 AU extending out to 1000 AU and possibly having non-Keplerian motions, while 1.4 mm continuum observations indicate a disk with an inner radius of 110 AU and an outer radius of 350 AU \citep [Pi\'{e}tu et al. 2005; see also][]{man97, cor05}. Near-IR interferometric studies resolve the inner 0.7 AU region of the disk \citep{mil01}.  However, previous mid-IR studies present somewhat contradictory results.  Marsh et al. (1995) report extended structure at 17.9 $\mu$m with a semi-major axis of 80$\pm$20 AU and an inclination of 75$^\circ$.  Chen \& Jura (2003), using Keck, do not confirm that detection of extended structure, and at 20.5 $\mu$m using deconvolved images with a resolution of 0.$\arcsec$6, Pantin et al. (2005) report an elliptical ring-like structure at an average distance of 280 AU from the star. In addition, Liu et al. (2004) resolve the inner disk interferometrically at 10.3 $\mu$m determining a size of 27$\pm$3 AU and an inclination of 45$^\circ$. There is also recent evidence that AB Aur could be the brighter component of a binary system, with a companion separation most likely between 1 and 3 arcseconds \citep{bai06}.

In this paper we present deep mid-infrared images of AB Aur obtained at Gemini North.  We have resolved the emission close to the star at 11.6 and 18.5 $\mu$m, and we find an additional extended component that appears to be roughly circularly symmetric.  We show how these observations of the thermal emission from dust in the AB Aur system help establish a more coherent picture of the dust geometry consistent with most observations at other wavelengths.

\section{Observations and Data Reduction}

Observations of AB Aur were obtained on 2003 November 7 using Michelle \citep{roc04} at Gemini North as part of an imaging survey of Herbig Ae/Be stars to be presented elsewhere (Mari\~{n}as et al., in preparation). Images (Fig. 1) were taken using the Si-5 filter (11.6 $\mu$m, $\Delta $$\lambda $=1.1 $\mu $m) and Qa filter (18.5 $\mu$m, $\Delta $$\lambda $=1.6 $\mu $m). The standard chop-and-nod technique was used to remove thermal background from the sky and the telescope. The total on-source time was 645 seconds for each filter.  The nearby point-spread-function (PSF) star PPM 94262 was observed before and after each observation of AB Aur.  The average values of the full-width at half-maximum (FWHM) intensity of the PSF star were 0.$\arcsec$28 at 11.6 $\mu$m and 0.$\arcsec$44 at 18.5 $\mu$m, comparable to the diffraction limits ($\lambda$/D) of the telescope at these wavelengths.  A Moffat function was fitted to the radial profile of each source to derive the value of the FWHM.  All observations were made at airmasses less than 1.2.

Using the standard stars Sirius and Vega for flux calibration and airmass correction, we derive flux densities for AB Aur of 21$\pm$2 Jy at 11.6 $\mu$m and 36$\pm$4 Jy at 18.5 $\mu$m.  These values fall within the spread of previous measurements for AB Aur, which is thought to be variable in the mid-IR.  In addition to the emission from dust, the measured fluxes contain mid-IR radiation from the stellar photosphere.  We estimated the contribution from the photosphere by using the NextGen models developed by Hauschildt et al.(1999), assuming a stellar effective temperature of 9500 K \citep{anc98}, log g = 4.0, and solar metallicity.  We scaled the model using UBV fluxes for AB Aur \citep{mal98}.  From this procedure, we estimate the photospheric flux densities to be 0.054 Jy at 11.6 $\mu$m and 0.021 Jy at 18.5 $\mu$m, which are negligible compared to the emission from dust.

\section{Source Size}

We show in Fig. 1 the contour levels for AB Aur and the PSF star scaled so that peak fluxes and the lowest contours are at the same flux level for both images.  The lowest contour represents the 3$\sigma$ flux level (three times the background noise) for the AB Aur images.  Figure 2 shows the values of the FWHM for AB Aur and the PSF star.  We clearly resolve a bright, compact inner emission region near AB Aur. Since integration time can influence the FWHM of the image due to effects such as guiding errors, rotation of the pupil and seeing (Radomski et al. 2006, in preparation), we divided our data into subsets of equal integration time for AB Aur and PSF star.  By comparing these subsets, we conclude that the difference of the FWHM for AB Aur and the PSF star is real and not an artifact of integration time.  Quadratic subtraction of the average FWHM values of the source and the PSF star gives source sizes for the strong compact emission of 17$\pm$4 AU at 11.6 $\mu$m and 22$\pm$5 AU at 18.5 \footnote{Quadratic subtraction to determine the intrinsic source FWHM applies strictly to the deconvolution of two gaussian distributions, for which the square of the convolved FWHM equals the sum of the squares of the FWHM of the two gaussian functions.  For slightly resolved sources observed near the diffraction limit, quadratic subtraction provides a reasonable estimate of the intrinsic source size}. Error bars in these measurements are based on the dispersion of the average FWHM values for AB Aur and PSF as seen in Fig. 2. The size of the compact emission is consistent with that determined by Liu et al. (2004) at 10 $\mu$m.

The right-hand panels in Fig. 1 show the residuals after substraction of the normalized PSF from AB Aur. This residual corresponds to fainter emission extending out to 2 arcsec (280 AU) from the star at 11.6 $\mu$m and to 2.5 arcsec (350 AU) at 18.5 $\mu$m.  This latter value equals the outer boundary of the dust emission seen in the millimeter continuum \cite{pie05}.  We do not see evidence in our images of a ring-like structure as proposed by Pantin et al. (2005).  At these wavelengths a fit of ellipses to the contour levels of AB Aur at radii between 0.$\arcsec$2 and 2.$\arcsec$0 indicates an approximately constant position angle of 80$^\circ$$\pm$11$^\circ$, with an inclination (under the assumption of flat disk geometry) i = $\cos^{-1}(\Delta y)/(\Delta x)$ of 29$^\circ$$\pm$11$^\circ$ at 11.6 $\mu$m and 12$^\circ$$\pm$12$^\circ$ at 18.5 $\mu$m, where $\Delta$y and $\Delta$x are the semi-major and semi-minor axis of the fitted ellipse.  The uncertainties in the measurements arise from multiple fittings as we vary the radii of the ellipse. This low, nearly face-on, inclination is consistent with recent optical and near-IR results \citep{gra99,fuk04}

\section{Dust Properties} \label{bozomh}

Even elementary considerations imply that the AB Aur disk is structurally complex.  In this section we show that radial variations probably exist in the grain size and/or composition, and that the extended mid-IR emission originates in an optically thin region bounding an optically thick layer near the disk mid-plane.

\subsection{Particles Temperatures and Sizes}

Because the most extended mid-IR emission is faint, we divide our images into three annuli centered on the star, each with width $\Delta$r = 100 AU.  This is equivalent to 2.3 and 1.6 times the resolution elements at 11.6 $\mu$m and 18.5 $\mu$m, respectively.  By assuming that the mid-IR emission is optically thin and that the measured fluxes at both wavelengths originate within the same region, we then calculate average color temperatures of 215$\pm$3 K, 189$\pm$6 K, and 184$\pm$10 K for the circumstellar dust within each of the three regions, with the highest temperature corresponding to the region nearest the star. The quoted uncertainties in these temperatures are due only to measurement errors in the flux densities, which are the relevant uncertainties when examining the radial trends in temperature and corresponding dust properties. The expected blackbody temperatures at these distances are 103 K at 50 AU, 59 K at 150 AU, and 46 K at 250 AU, which are much lower than the derived color temperatures, indicating the presence of smaller, less efficiently emitting grains.

The temperature of a dust grain in thermal equilibrium with stellar radiation depends on the distance to the heating source and on the radiative efficiency of the dust particle.  The value of the efficiency depends on the properties of the material and the size of the grain and can be considered equal to unity for radiation shorter than a critical wavelength $\lambda_o$.  Given some assumptions about the material, the value of $\lambda_o$ can be used to estimate a characteristic grain radius $a$.  One finds that $\lambda_o \approx a$ in the case of moderately absorbing dielectrics like graphite and amorphous silicate, while $\lambda_o/2\pi\approx a$ for strongly absorbing dielectrics like polycyclic aromatic hydrocarbons (PAHs) \citep{bac93}. We used the energy balance equations in the form presented in Backman and Paresce (1993) to calculate the expected flux density ratio (11.6/18.5) for different values of $\lambda_o$ as a function of distance to the star and constrain the sizes of the mid-IR emitting particles throughout the disk. From our observations, the observed flux ratios for the three regions are 0.63-0.68, 0.46-0.53, and 0.44-0.51, with the values decreasing as we move further from the star. Assigning these average flux ratio values to the midpoint of each region and assuming moderately absorbing materials, we see that the emission in these regions can be well constrained to come from particles of sizes 1.1 to 1.3 $\mu$m in the central 100 AU region, 0.2 to 0.3 $\mu$m for the dust between 100 and 200 AU, and 0.08 to 0.12 $\mu$m for the dust between 200 and 300 AU, see Fig 3.  For strongly absorbing material, the corresponding average values are 0.2 $\mu$m for the compact emission and 0.04 $\mu$m and 0.01 $\mu$m for the other two regions.

These considerations suggest that we need different dust populations to fit the color temperatures at different distances from the star.  The circumstellar dust in AB Aur is probably a combination of different materials, since PAH emission bands at 3.4, 6.2, 7.7, 8.6, and 11.3 $\mu$m and the amorphous silicate emission feature at 9.7 $\mu$m have been detected \citep{coh80, anc00}.  Thus, the deduced radial variation in particle size may result partly or entirely from a radial variation in dust composition. Pi\'{e}tu et al. (2005) find that the CO emission in AB Aur extends out to 1000 AU, whereas the millimeter dust continuum, like out 18 $\mu$m emission, extends only to 350 AU.  They propose that this marked difference in the CO and dust continuum radial distributions may be due to a fairly abrupt change in the radial variation in the dust opacity, perhaps associated with less-processed dust at larger radii.  Thus, at least qualitatively, both sets of observations support the idea that there are radial variations in the dust properties  in the AB Aur disk.  Mid-IR spectroscopy from the VLTI of three other Herbig Ae stars has also revealed radial variations of dust composition in those systems, with crystalline silicates dominating the inner 2 AU region, and a mixture of crystalline and amorphous silicates located in the 2-20 AU region \citep{van04}. Detailed follow-up mid-IR spectroscopy with high spatial resolution can help resolve this issue.

\subsection{Dust Optical Depth and Disk Morphology}

Stellar optical radiation penetrates the CS dust to a radius corresponding roughly to $\tau_\nu$ = 1.  We assume that the disk is flat, with an inner radius R$_{in}$ = 0.5 AU, an outer radius R$_{out}$ = 400 AU, a thickness $\Delta$S = 10 AU, and a dust mass M$_{mm}$ = 2 x 10$^{29}$ g \citep{man97}.  The resultant uniform volume dust density is given by the relationship,

\begin{equation}
\rho_o = \frac{M_{mm}}{\pi(R_{out}^2 - R_{in}^2)\Delta S}
\end{equation}

The optical depth along the plane of the disk is

\begin{equation}
\tau_\lambda = \kappa_\lambda \int_{R_{in}}^{R}\rho_o dr
\end{equation}

with the absorption coefficient of the form $\kappa_\lambda = \kappa_{\lambda o}(0.25 mm)/\lambda$, where $\kappa_{\lambda o}$ = 0.1 $cm^2 g^{-1}$ \citep{hil83}.

We estimate from these considerations that $\tau_\nu$ =1 at R $\sim$ 2 AU.  The value obtained is an upper limit for several reasons. First, for simplicity we have assumed a uniform volume dust density; however, a radially decreasing density distribution with power-law indices in the range 0.5 to 2 is more consistent with models \citep{bec90, men97, dul01} and will increase the dust density closer to the star.  Second, extrapolating values for the visible absorption coefficient from the submillimeter region underestimates the visible extinction along the line of sight, which is five times larger than predicted by this relationship at 0.55 $\mu $m \citep{mat90}; therefore, the value of $\tau_\nu$ = 1 is probably reached at R $<$ 2 AU. We conclude that the stellar radiation does not penetrate very far in the plane of the disk, in agreement with the conclusion of Mannings \& Sargent (1997).  However, for the dust far from the star to reach the temperatures inferred from our observations, it must be heated by direct radiation from the star, which implies heating of the dust above and below the mid-plane. This dust could reside either in the surface layer of a flaring disk or in an envelope.

To address this issue further, we consider the vertical (i.e. perpendicular to the plane of the disk) optical depth in the system.  Since we believe the disk to be almost face-on, an estimate of the vertical optical depth of the disk at mid-IR wavelengths should indicate if we are only detecting emission from the dust located above the disk (if the disk is optically thick to mid-IR radiation) or from both sides of the disk (if the disk is optically thin to mid-IR radiation).  The dust in the mid-plane evidently would be too cool for us to detect, since visual radiation cannot penetrate there.  Using the best fit to the AB Aur data for the disk surface density distribution $\Sigma = \Sigma_o (1 AU/r)^{2}$, where $\Sigma_o = 10^4$ g cm$^{-2}$ \citep{dul01} and $\tau_\lambda = \kappa_\lambda \Sigma$, we find that the AB Aur disk is optically thick vertically to mid-IR radiation out to a radius of about 118 AU.  The disk becomes optically thin to mid-IR radiation beyond 118 AU from the star.  This result suggests that, except in the outermost disk, we cannot look through the disk at mid-IR wavelengths and are only detecting mid-IR emission from the "surface" layer of the disk facing us. This result has its parallel in the CO observations of AB Aur by Pi\'{e}tu et al. (2005), which show that the optically thick emission lines, arising near the more directly irradiated disk surface, have higher excitation temperatures than the optically thin ones weighted toward material in the disk mid-plane.

\subsection{Models}

The passive flared disk model with an inner rim developed by Dullemond, Dominik, \& Natta (2001) returns the detailed geometry of a CS disk for given specific stellar parameters and general disk properties such as dust mass and inner and outer disk radii.  We have applied their model code (kindly provided by C. P. Dullemond) to AB Aur to compare the inferred geometry of the disk model to parameters derived from our observations. In this context, the size of the compact mid-IR emission detected in our images coincides with the emergence at a radius of $\sim$ 10 AU of the disk from the shadow of the inner rim.  The model predicts a jump in the surface temperature of the dust at this boundary, which translates into stronger fluxes at mid-IR wavelengths.  The dust temperature implied by our images for the compact mid-IR emission is about 100 K lower than the value predicted by the model for the surface layer of the disk at the onset of flaring (see Fig. 3 in Dullemond et al. 2001). This inconsistency results from the fact that we are comparing an average color temperature within the inner 100 AU to the peak temperature from the model within this region.  With the exception of the dust at the inner rim, which is heated to dust sublimation temperatures, the modeled dust temperatures in the inner 10 AU are colder than 200 K because dust in that region is shadowed from stellar radiation.  The dust temperature peaks at a radius of about 10 AU and decreases outwards as r$^{-2}$.  The model indicates that the average temperature of the surface layer in this region is about 200 K, which is consistent with our results in the previous section.  In this context, we suggest that the strong compact emission detected in our images is a combination of unresolved emission from the inner rim, and emission from the surface layer at the onset of flaring, which is resolved. Likewise, the size of the more extended component that we detect in our images is as predicted from the model for the outer surface layer of the disk, which becomes too cold at very large radii to be detected in the mid-IR.

The segregation of particle sizes that we derive from our observations can also be explained within the passive flared disk model.  Small grains in a region optically thin to stellar radiation, like the surface layer of the disk, experience the effect of radiation pressure.  The parameter $\beta$ is the ratio of the radiative to gravitational forces, which is proportional to the luminosity of the star and inversely proportional to the particle size. For $\beta > 0.5$, particles are on unbound orbits and will be expelled from the system in timescales comparable to the orbital period of the region where the particles were produced \citep{bac93}.  In the case of AB Aur, these particles will be expelled in timescales of less than 10$^{3}
$ years for a 500 AU disk in a gas-free environment. We do not consider the presence of molecular CO at radial distances comparable to the mid-IR emission as relevant for this result because we believe that the gas and the mid-IR emitting particles occupy different vertical regions of the disk.  When gas molecules collide frequently with dust particles, as is the case when abundant gas and dust co-exist, they reach similar temperatures; gas temperatures can drop below dust temperatures as gas is depleted in a system.  Pietu et al. (2005) derive temperatures of 70 K for the warm gas near the surface of the disk at a radius of 100 AU and found no evidence of CO depletion.  Our derived dust temperatures for this radial distance are substantially higher; therefore, it is appropriate to assume that the mid-IR emitting dust resides in an optically thin layer at the surface of the disk, while the warm gas detected in the system lies at a lower elevation above the mid-plane. Even in the case that gas and dust co-exist, Klahr \& Lin (2000) showed that hydrodynamic drag forces in a gas rich environment tend to enhance the effect of radiation pressure, increasing the critical size for particles blown out of the system in a dynamical timescale.  The presence of micron-size particles in a system 2-4 Myr old implies that there is a replenishing mechanism for these grains, probably collisions of larger particles \citep{wya99}.  Regardless of the formation process involved, if small particles are constantly created, then we should see them at those locations where they are produced or as they are driven outward by radiation pressure from regions closer to the star. They should have lower emission efficiencies, and therefore, elevated temperatures compared to larger dust at the same distance from the star.  These smaller, hotter particles would be easier to detect in the mid-IR farther from the star.

We also considered the implications of our observations if we were detecting emission from a disk and a halo, as proposed by Vinkovi\'{c} et al. (2003).  The presence of a halo could explain the difference between the derived inclination angles from our study and that of Liu et al. (2004).  Given that our images achieved higher sensitivity than those of Liu et al. (2004), we could be seeing more of the tenuous halo or spherical component, which will dilute the asymmetries created by an inclined disk.  However, the variation of grain sizes derived from our observations is inconsistent with emission from a halo.  In the roughest approximation, if a halo were present we would expect to find a population of larger colder grains concentrated in the disk, and smaller micron and sub-micron size particles residing in the halo.  If this were the case, the average grain size in the inner region should be smaller than at larger distances because we would be intercepting more of the halo material in the central region, which is the opposite of our findings.

\section{Conclusions}

Our mid-IR images reveal two different emission components in AB Aur.  The central stronger emission is resolved, with quadratically deconvolved FWHM sizes of 17$\pm$4 AU and 22$\pm$5 AU at 11.6 $\mu$m and 18.5 $\mu$m, respectively.  We also detect fainter extended emission out to a radius of 280 AU at 11.6 $\mu$m and 350 AU at 18.5 $\mu$m.  Emission is slightly elongated at 12 $\mu$m, indicating a disk inclination angle in the range 29$^\circ \pm 11^\circ$ and a PA of 80$^\circ \pm 11^\circ$.  The morphology at 18 $\mu$m is consistent with an inclination angle of 12$^\circ \pm 12^\circ$. However, within the uncertainties, inclination angles at 12 and 18 $\mu$m are the same.  Assuming moderately absorbing material, we derive average radii of the mid-IR emitting dust in the system and find that larger particles ($a \sim 1$ $\mu$m) dominate the mid-IR emission in the inner ($<$ 100 AU) regions of the disk, and smaller particles ($a <$ 0.3 $\mu$m) dominate in the outer regions of the disk.  Our results are reasonably well accounted for by a model of a passive flared disk with an inner rim.  The presence of a more spherical component fails to account for the particle size segregation derived from our observations.

\acknowledgments

We are grateful to C. P. Dullemond for providing his model code and to the referee for his comments, which help clarify and improve this paper. NM acknowledges the NASA Graduate Student Research Program for financial support (NASA Grant NGT5-50473 to CMT). This research was also supported in part by NSF grant AST-0098392 to CMT and by NSF grant AST-0206617 to CP.

{\it Facility:} \facility{Gemini:Gillett (Michelle)}

\clearpage



\begin{figure}
\epsscale{0.8}
\plotone{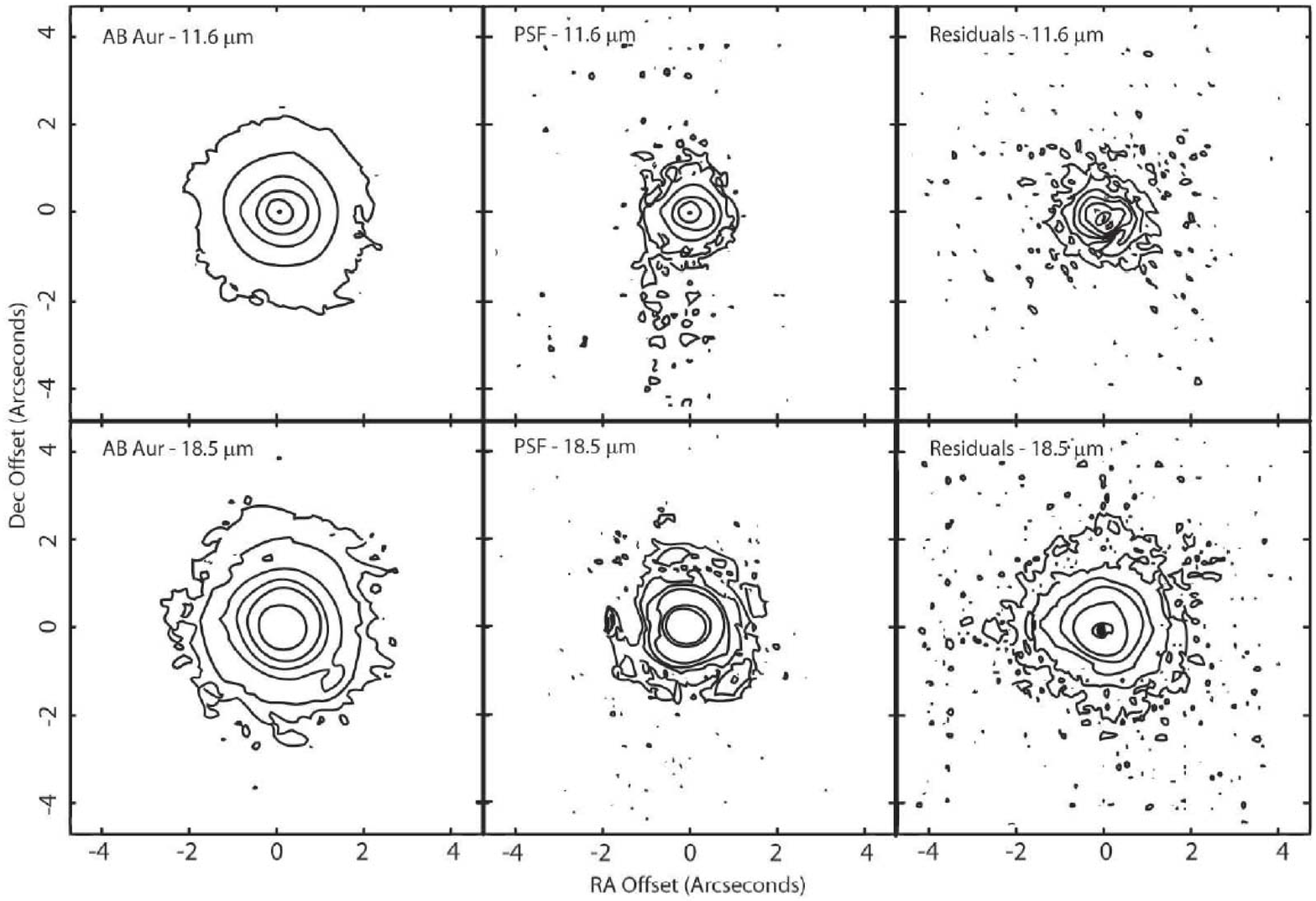}
\caption{Upper panels show 11.6 $\mu$m image (contour) of AB Aur, the PSF star scaled to 100 \% of AB Aur peak emission, and the PSF subtracted emission. The AB Aur and PSF star contour levels are (0.06, 0.25, 1.07, 4.50, 18.86, 79.20) Jy/arcsec$^{2}$, lowest contour is 3 $\sigma$ (60 mJy/arcsec$^{2}$) for AB Aur.  The contour levels for the residual emission at 11.6 $\mu$m are (0.37, 0.69, 1.31, 2.47, 4.67, 8.83) Jy/arcsec$^{2}$.  Lower panels show the 18.5 $\mu$m image (contour) of AB Aur, the PSF star scaled to 100 \% of AB Aur peak emission (Sirius was used to obtain better signal-to-noise ratio at this wavelength), and the PSF subtracted emission. Contour levels are (0.15, 0.33, 0.73, 1.61, 3.54, 7.77) Jy/arcsec$^{2}$. The lowest contour is 3$\sigma$ (150 mJy/arcsec$^{2}$).  Emission from AB Aur can be seen extending to $\sim2$ arcsec in the 11.6 $\mu$m data and to $\sim2.5$ arcsec in the 18.5 $\mu$m data.  For cosmetic reasons, a very noisy part of the image at the extreme right of the 11.6 $\mu$m PSF image has been removed; that has no effect in the region of the stellar image.\label{fig1}}

\end{figure}

\clearpage

\begin{figure}
\epsscale{0.5}
\plotone{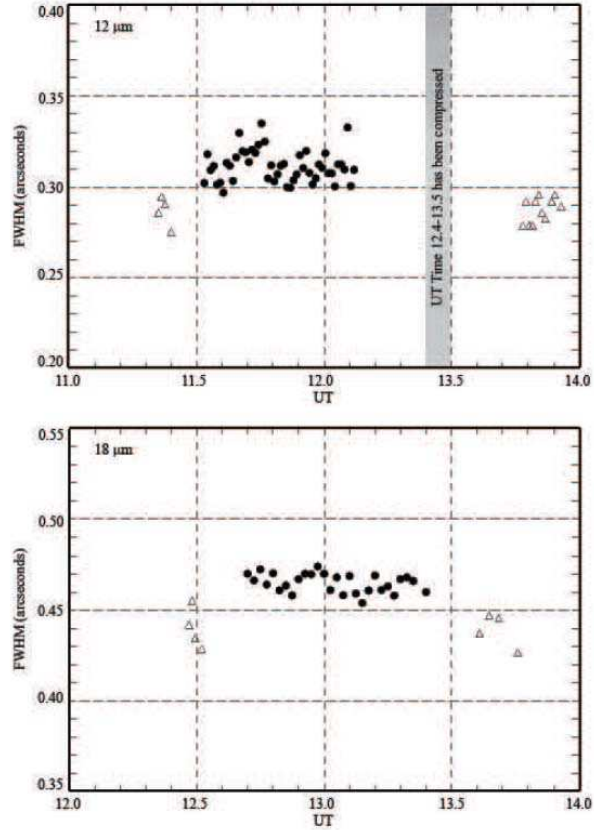}
\caption{This figure shows 11.6 $\mu$m (top panel) and 18.5 $\mu$m (lower panel) FWHM values for AB Aur (solid circles) and PSF star (open triangles).  Each point of the 11.6 $\mu$m data represents the signal level of 1 nod position for a total on-source time of 13.44 sec. Each point of the 18.5 $\mu$m data represents the signal level of 2 nod positions stacked for a total on-source time of 23.04 sec. After the 11.6 $\mu$m observations of AB Aur were taken, the PSF star was going through transit very close to zenith, as a result, there is a large elapse time in the graph between the observations. The second PSF star observations at 11.6 $\mu$m was taken at the end of the 18.5 $\mu$m observations.\label{fig2}}
\end{figure}

\clearpage

\begin{figure}
\epsscale{1.0}
\plotone{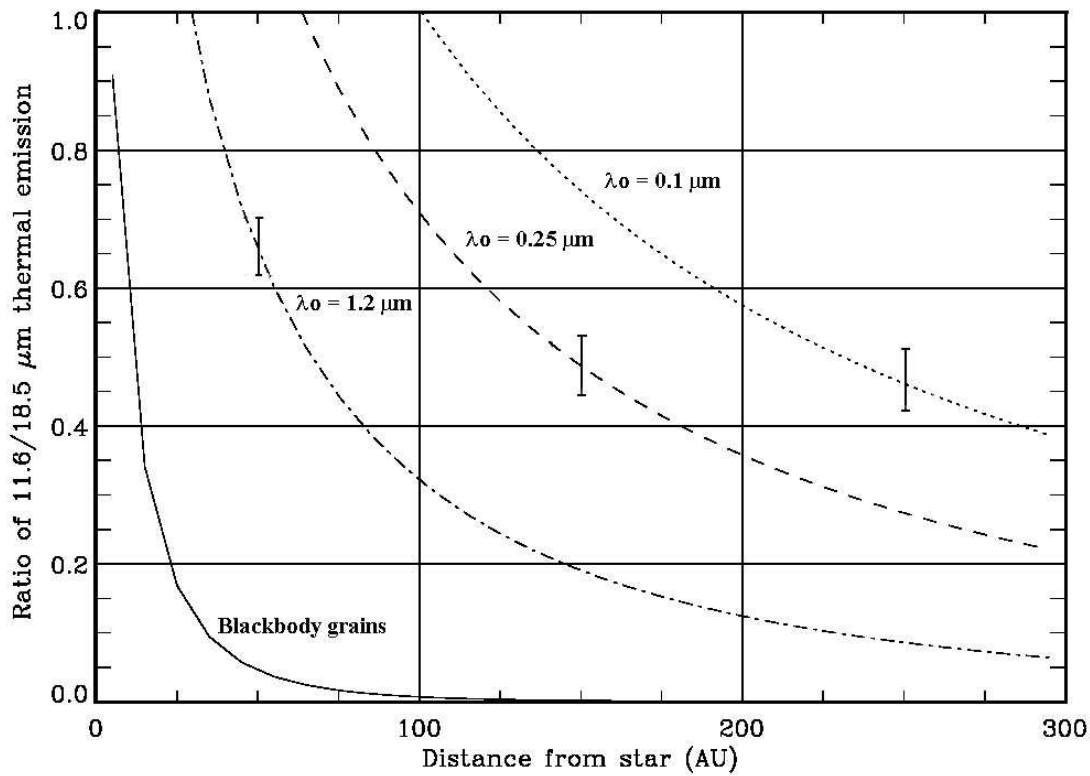}
\caption{The ratio of the thermal emission at 11.6 $\mu$m and 18.5 $\mu$m is shown for different values of $\lambda_o$. Grain sizes equal $\lambda_o$ for moderately absorbing material or $\lambda_o/2\pi$ for strongly absorbing materials \citep{bac93}. The observed ratio of flux densities for the regions 0 to 100 AU, 100 to 200 AU, and 200 to 300 AU are shown at the mid-point of the three regions, at 50, 150, and 250 AU from the star. The error bars represent the relative uncertainties in the flux measurement ratios.\label{fig3}}
\end{figure}

\end{document}